\documentclass[12pt]{article}
\pdfoutput=1
\usepackage{putex}
\usepackage{graphicx}
\usepackage{enumerate}
\usepackage{cite}
\usepackage{tensor}
\usepackage{slashed}
\usepackage{hyperref}
\usepackage{float}
\usepackage[bottom]{footmisc}
\usepackage[utf8]{inputenc}
\usepackage{amsmath}
\usepackage[usenames,dvipsnames,svgnames,table]{xcolor}
\usepackage[disable]{todonotes}

\usepackage[compat=1.1.0]{tikz-feynman} 
\usetikzlibrary{positioning}
\usetikzlibrary{decorations.pathmorphing}
\usepackage{tikz-feynhand}

\hypersetup{
	pdfstartview={FitH},    
	pdftitle={Hamiltonian approach to near-extremal black hole evaporation and backreaction},    
	pdfauthor={Kraus},     
	colorlinks=true,       
	linkcolor=blue,          
	citecolor=blue!59!black! ,        
	filecolor=magenta,      
	urlcolor=blue!75!black,           
	linktoc=all
}

\makeatletter
\g@addto@macro\bfseries{\boldmath}
\makeatother

\numberwithin{equation}{section}

\newcommand{\eq}[2]{\begin{align}\label{#1}#2\end{align}}

\newcommand {\be} {\begin {equation}}
\newcommand {\ee} {\end {equation}}

\newcommand{\p}{\partial}

\newcommand\rt{{\rightarrow}}
\def\eps{\epsilon}

\newcommand{\rf}[1]{(\ref{#1})}
\newcommand{\rff}[1]{\ref{#1}}

\newcommand{\phit}{\tilde{\phi}}
\newcommand{\veps}{\varepsilon}
\newcommand{\ut}{\tilde{u}}
\newcommand{\vt}{\tilde{v}}

\definecolor{greenC}{rgb}{0.0, 0.38, 0.18}

\newcommand{\Tt}{\tilde{T}}
\newcommand{\Hc}{{\cal H}}
\newcommand{\Mc}{{\cal M}}

\begin{document}

	\institution{UCLA}{ \quad\quad\quad\quad\quad\quad\quad\ ~ \, $^{1}$Mani L. Bhaumik Institute for Theoretical Physics
		\cr Department of Physics \& Astronomy,\,University of California,\,Los Angeles,\,CA\,90095,\,USA}

	\title{Hamiltonian approach to backreaction in near-extremal black hole evaporation
	}
	
	\authors{Per Kraus$^{1}$}
	
	\abstract{We investigate radiation from near-extremal black holes formed by collapse, focusing on the role of large backreaction effects arising from gravitational fluctuations in the near-horizon region. Such effects have previously been identified from computations based on JT gravity and its Schwarzian description, most notably for the Euclidean partition function.   Restricting attention to the s-wave sector, we integrate out gravity by solving the constraint equations in the Hamiltonian formalism, obtaining an effective scalar action with a coupling that grows at low temperature, thus enabling a real-time treatment of quantum backreaction. We then take initial steps toward evaluating the impact of this interaction on correlations of the outgoing radiation, and compare our findings with earlier results.	
 }
	
	\date{}
	
	\maketitle
	\setcounter{tocdepth}{2}
	\begingroup
	\hypersetup{linkcolor=black}
	\tableofcontents
	\endgroup
	

\section{Introduction}

Near-extremal black holes have proven to be a fruitful playground for the study of qualitatively important quantum gravity  effects in  a regime  where  explicit computations are possible despite  incomplete knowledge of the underlying theory.   Even though the spacetime curvature near the event horizon of a large near-extremal black hole is small, there are nonetheless large fluctuations of the gravitational field, a fact inferred from thermodynamic considerations in \cite{Preskill:1991tb}, and developed in much detail following the study  of  backreaction  in  potential AdS$_2$/CFT$_1$ duals  \cite{Almheiri:2014cka}.   The identification of  JT gravity \cite{Jackiw:1984je,Teitelboim:1983ux} as capturing near-horizon physics \cite{Almheiri:2014cka}, and its description in terms of Schwarzian quantum mechanics \cite{Jensen:2016pah,Maldacena:2016upp,Engelsoy:2016xyb,Stanford:2017thb} led to a host of new results, including a computation of the density of states of near-extremal black holes, exhibiting a dramatic departure from the Bekenstein-Hawking area law formula \cite{Iliesiu:2020qvm,Heydeman:2020hhw}.  See \cite{Mertens:2022irh} for a review.   This departure is driven by the fact that the expansion parameter governing gravitational interactions near the horizon is not the naive (in 4d)   ${G\over r_0^2}$, with  $r_0$ being the horizon size, but is instead enhanced at low temperature to ${G\over r_0^3 T_H}$.   This enhancement effect is visible by examining  the action for metric fluctuations near the horizon \cite{Almheiri:2016fws} and is of the correct size to account for the thermodynamic breakdown discussed in \cite{Preskill:1991tb}.   Further elucidation of these large quantum gravity effects would be beneficial.

Our objective here is to capture the effect of large gravitational fluctuations in essentially the same setup as in Hawking's original computation  \cite{Hawking:1975vcx}:  start in the vacuum in the far past, then form a black hole by collapse while evolving the quantum field forward in time to see what comes out.  To make the problem tractable we restrict to spherical symmetry and model black hole formation by collapse of a charged null shell.  Before any matter or quantum effects are taken into account this yields the spacetime depicted in fig. \rff{collapse}, corresponding to the formation of a 4d  Reissner-Nordstr\"om black hole of  mass $M$ and charge $Q$.
\begin{figure}[H]
\centering
\includegraphics[scale=0.349]{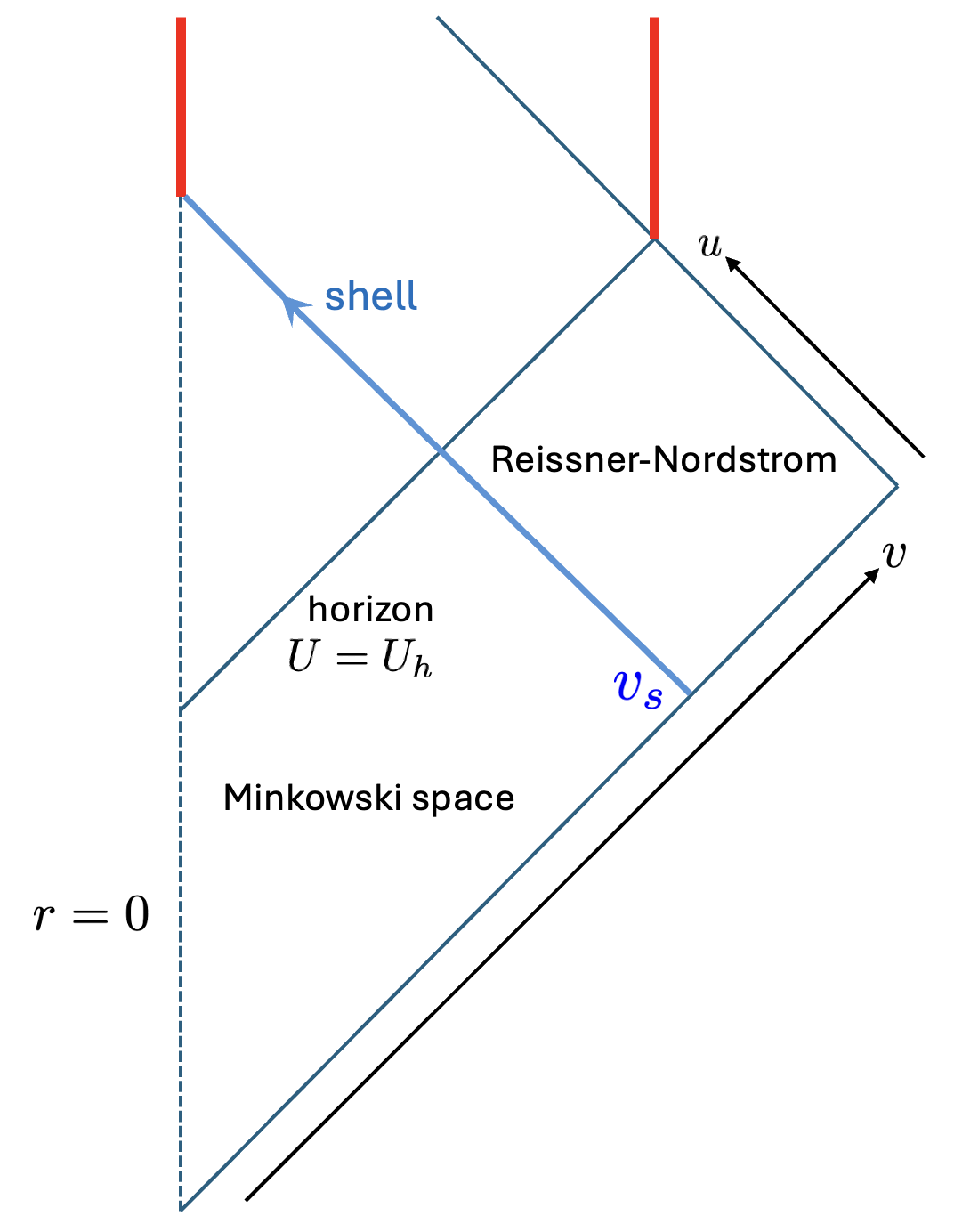}
\caption{ Collapse of a  charged null shell forming a Reissner-Nordstr\"om black hole.        }
\label{collapse}
\end{figure}
We aim to compute  in-in expectation values of scalar field operators at future null infinity.  To probe the expected ${G\over r_0^3 T_H}$ effects, the temporal separation between scalar field operator insertions is taken to be of order the thermal wavelength.  As noted above, the effects of interest come from large gravitational fluctuations near the horizon, so we need a way to incorporate these into the usual derivation of Hawking radiation from collapse.  We proceed according to the  strategy outlined below, but first pause to comment on some work that bears on related issues.

The cleanest arena for the study of gravitational fluctuations is the Euclidean path integral for JT gravity in a (regularized) AdS$_2$ geometry.  On the other hand, our problem is inherently Lorentzian and involves asymptotically flat spacetime.  Regarding the (Euclidean)  AdS$_2$ versus flat space asymptotics, one approach, employed in \cite{Iliesiu:2020qvm}, is to consider the full asymptotically flat spacetime and then effectively factor the problem by a matching procedure at the overlap  of the near-horizon  and asymptotic regions.  At the level of the one-loop fluctuation determinant, it is also possible to isolate the modes responsible for the low temperature enhancement directly in the full asymptotically flat geometry \cite{Kolanowski:2024zrq}, at least numerically.   Classically, the emergence of the JT description can be studied by computing boundary correlation functions of a near-extremal black hole with AdS$_{d>2}$ asymptotics \cite{Nayak:2018qej,Moitra:2018jqs}.   Regarding real-time dynamics and the emission and absorption of radiation,  \cite{Brown:2024ajk,Lin:2025wof,Emparan:2025sao,Biggs:2025nzs,Emparan:2025qqf,Betzios:2025sct} have taken the approach of using the Euclidean Schwarzian theory to extract coarse-grained matrix elements of field operators between black hole states, and then used these to couple the black hole to an external quantum field, much as in the description of photon emission/absorption from atoms.

Returning to our problem, to follow the same path as in Hawking's  computation  we need an efficient method to include the important gravitational fluctuations.  To this end, we derive a gravitationally dressed action for the scalar field.
Restricted to spherical symmetry, gravity has no independent local degrees of freedom and hence can be integrated out to yield an effective action for the matter, here taken to be a real scalar field.   In order to arrive at an as-local-as-possible effective action, one should integrate out by solving the canonical gravitational constraint equations.   This yields an action which is nonlocal in space but local in time, and hence can be treated by conventional Hamiltonian methods at both classical and quantum levels.  Such an effective action was originally written down in \cite{Berger:1972pg}, with a corrected version provided in \cite{Unruh:1976db}.  We work out a generalization to include charge and the presence of the collapsing shell.  With an appropriate choice of gauge, the  classical Hamiltonian can be found explicitly, 
%
%
%
\eq{A1}{ H_{\rm ADM} & =   M + \int_{r_s}^\infty \! dr' f(r') h(r')  e^{-\int_{r'}^\infty \! dr'' {2Gh(r'')\over r''}}~.  }
Here $M$ and $Q$ represent the mass and charge of the shell, $r_s$ is the shell radius, $f(r)=1-{2GM\over r}+{GQ^2 \over r^2}$ is the function appearing in the Reissner-Nordstr\"om metric,  and the energy density of the scalar field is represented by $h(r)$,
\eq{A2}{  h(r) = {1\over 2} \left( {\pi_\phi^2 \over r^2} +r^2 \phi'^2\right)~.}
The gravitationally dressed scalar  action is then $S= \int\! dt dr \pi_\phi \dot{\phi} - \int\! dt H_{\rm ADM}$.  Obviously,  quantization based on this full action is bound to be subtle, requiring UV counterterms and a careful effective field theory analysis of which IR observables are calculable.  This will not be attempted here, where we instead work to first order in gravitational perturbation theory, corresponding to expanding the action to quartic order in the scalar field.  Our interest is in understanding how this action leads to enhanced $G/T_H$ effects in the presence of a near-extremal black hole.   To study this, we implement a standard near-extremal, near-horizon scaling of the parameters, and work out the form of the quartic interactions in the near-horizon region.   This leads to the following  gravitationally dressed  action governing the near-horizon region,
\eq{A3}{  S& = \int\! d\tau \Bigg[ 2 \int_{\rho_s}^\infty {d\rho\over \rho^2-\rho_h^2} \p_{\ut} \phit \p_{\vt} \phit \cr
& \quad +{\rho_h G\over  \pi  r^3_0 T_H} \int_{\rho_s}^\infty {d\rho \over \rho^2- \rho_h^2}  \int_{\rho}^\infty {d\rho' \over \rho'^2-\rho_h^2} { \big[ (\p_{\ut} \phit)^2+(\p_{\vt} \phit)^2 \big]\big[ (\p_{\ut'} \phit)^2+(\p_{\vt'} \phit)^2 \big]\over \rho'^2-\rho_h^2   }  \Bigg]~,}
where the top line is the free scalar action in the AdS$_2$ black hole metric
$ ds^2  = -(\rho^2 -\rho_h^2)d\tau^2 + {d\rho^2 \over \rho^2 -\rho_h^2}$ or in light cone coordinates,  $  ds^2     = - (\rho^2 -\rho_h^2)d\ut d\vt,$
and where $\phit= \phi/r_0$.   The key point to note here is that the coefficient of the interaction is proportional to ${G\over r_0^3 T_H}$, which is the sought after near-extremal enhanced coupling.

Our working assumption to capture the enhanced effects is that we can use this action to compute correlation functions in the near-horizon region, which subsequently can be propagated out to asymptotically flat infinity using the free field equations.  In this way we can study scattering processes and gravitational correlations in the Hawking radiation.  In this paper we just take an initial step in this direction by examining (part of) the one-loop correction to the 2-point function.   An analogous quantity was worked out in \cite{Maldacena:2016upp}, and was found to grow (relative to the tree-level result) for large time separations as  $ {G\over r_0^3 T_H } \rho_h^2 \tau^2$, in our conventions. This growth was further given a simple physical explanation in terms of energy fluctuations of the black hole.  In our case, we also find such a term, however it is accompanied by an exponentially growing term proportional to $e^{\rho_h \tau}$.    As we discuss in section \rff{pert} the difference is that the computation in \cite{Maldacena:2016upp} corresponds to a black hole in thermal equilibrium, while we have outgoing radiation only, so that the black hole shrinks.  Including an incoming thermal flux to balance the outgoing radiation is expected to remove the exponential term by cancelling the tadpole diagram that supports it.

The remainder of this paper is organized as follows.
In section \rff{review} we review some basics, namely the Reissner-Nordstr\"om solution, the near-horizon scaling, the collapse geometry, and Hawking radiation in the free field limit.  In section \rff{BCMN} we work out the effective scalar action by integrating out gravity in the s-wave.  The perturbative computation of correlators is discussed in section \rff{pert}, and we conclude with a discussion in section \rff{discussion}.  Two appendices contain technical results.

\section{Gravitational collapse of a charged shell, and free field Hawking radiation}
\label{review}

This section reviews the spacetime resulting from the collapse of a spherical charged null shell, including  a discussion of the near-extremal, near-horizon limit, followed by a derivation of Hawking radiation from this spacetime in the free field approximation.  Essentially everything in this section is found in standard references like \cite{Birrell:1982ix}.

\subsection{Reissner-Nordstr\"om  solution}

The 4d asymptotically flat Reissner-Nordstr\"om black hole has the following line element in Schwarzschild coordinates
\eq{a1}{  ds^2 = -f(r)dt^2 + {dr^2\over f(r) } +r^2 d\Omega^2 }
with
\eq{a2}{ f(r)&= 1-{2GM\over r } +{GQ^2\over r^2} = {(r-r_+)(r-r_-) \over r^2}
\cr  r_\pm &= GM \pm \sqrt{G^2M^2 -GQ^2}~. }
The Hawking temperature, computed from the surface gravity at the event horizon $r=r_+$, is
\eq{a3}{ T_H ={ f'(r_+) \over 4\pi} = {r_+-r_-\over 4\pi r_+^2}~.}
The extremal limit is $r_+=r_-$, equivalently  $GM^2=Q^2$.

The light cone coordinates $(u,v)$ are defined as
\eq{a3a}{ u =t-r_*~,\quad v=t+r_*}
where the tortoise coordinate $r_*(r)$ is
\eq{a3b}{ {dr_* \over dr } = {1\over f(r)}~, }
so that the line element becomes
\eq{a3c}{ ds^2 = -f\big(r(r_*)\big) dudv + r^2(r_*) d\Omega^2~.}
Explicitly,
\eq{a4x}{ r_* =   r+ {r_+^2 \over r_+-r_-} \ln(r-r_+) -  {r_-^2 \over r_+-r_-} \ln(r-r_-) }
and the horizon $r=r_+$ is located at $r_*=-\infty$.

\subsection{Near-horizon limit}

We now review the near-extremal, near-horizon limit.  We define $r_0$ and $\eps$  by writing
\eq{a4}{ r_\pm = r_0 \pm \eps~.}
The coordinates $(t,r)$ are traded for $(\tau,\rho)$ as
\eq{a5}{ t=\frac{r_0^2}{\lambda} \tau~,\quad r=r_0+\lambda \rho  }
and we also redefine $\eps$, and introduce $\rho_h$,  as
\eq{a6}{ \eps = \lambda \rho_h~.}
The scaling parameter $\lambda$ controls the approach to the extremal limit.
The strict near-horizon limit is obtained by taking $\lambda \rt 0$ at fixed $(\tau,\rho,r_0,\rho_h)$, which yields the AdS$_2 \times S^2$ metric.  We instead think in terms of small but finite $\lambda$, focusing on the leading behavior as $\lambda \rt 0$,
\eq{a7}{ ds^2  = r_0^2 \left[-(\rho^2-\rho_h^2) d\tau^2 +{d\rho^2 \over \rho^2 -\rho_h^2} \right] +(r_0^2 +2\lambda r_0 \rho)d\Omega^2~.}
For illustration we have retained the leading correction to the sphere radius; in the 2d JT gravity description this represents a linearly growing dilaton, and in that context one fixes a cutoff surface by fixing the value of the dilaton on that surface.
In the extremal limit  $r_0= GM = \sqrt{GQ^2}$.  The temperature, as measured at asymptotically flat infinity, goes to zero in the extremal limit, with leading behavior
\eq{a8}{ T_H \approx {\lambda \rho_h \over 2\pi r_0^2}~.}
The scaling parameter $\lambda$ may thus be traded for the temperature $T_H$.

\subsection{Collapse geometry}

We adopt the simplest model of Reissner-Nordstr\"om black hole formation, namely the collapse of a charged spherical null shell.  The  spacetime is described by patching a region of Minkowski space inside the shell to a Reissner-Nordstr\"om solution outside the shell; see fig.  \rff{collapse}.

Writing the Minkowski metric as $ds^2 = -dT^2 + dr^2  + r^2 d\Omega^2 $ we define
\eq{b1}{ U = T-r~,\quad V=T+r}
so that inside the shell
\eq{b2}{ ds^2  =-dUdV + r^2(U,V)d\Omega^2~.}
The infalling spherical shell is taken to lie at $V=V_s$, so that the metric takes the form \rf{b2} for $V<V_s$. The radial trajectory of the shell can then be written as $r(U) = {V_s-U\over 2}$.

The metric exterior to the shell may be written as in \rf{a3c}, with the shell lying at $v=v_s$ and the exterior given by $v>v_s$.  The radial shell trajectory expressed in the exterior coordinates is  $r(u) = r\Big( r_*= \big({v_s-u\over 2}\big)\Big)$.

What remains is to relate the interior coordinates $(U,V)$ to the exterior coordinates $(u,v)$. Using our freedom to shift the time coordinate there is no loss in generality in taking $V=v$.   The relation between $U$ and $u$ is found by equating the physical shell radius $r$ in the two coordinates,
\eq{b3}{  {v_s-U\over 2} = r\Big( r_*= \big({v_s-u\over 2}\big)\Big)~.}
This gives
\eq{b4}{ u & = U-{2r_+^2 \over r_+-r_-} \ln\left({v_s-U\over 2}-r_+\right) +  {2r_-^2 \over r_+-r_-} \ln\left({v_s-U\over 2}-r_-\right)~. }
The event horizon at $u=\infty$ is mapped to the finite value
\eq{b4a}{ U=U_h\equiv  v_s -2r_+~.}
The region $U> U_h$ is inside the horizon.  We also record
\eq{b5}{ {du\over dU} = {(v_s-U)^2 \over (v_s-2r_+-U)(v_s-2r_--U)}~.}
While it is not possible  to invert \rf{b4} analytically to find $U(u)$, the late time ($u\rt \infty)$ behavior can be obtained as
\eq{b6}{ U = U_h - C e^{-2\pi T_H u} + \ldots~,\quad C =2 (r_+-r_-)^{r_-^2/r_+^2} e^{2\pi T_HU_h}~.}
This relation governs the form of the radiation at late times.

\subsection{Hawking radiation in the free field limit}

We now introduce a minimally coupled free scalar field,
\eq{c1}{ S = -{1\over 8\pi} \int\! d^4x \sqrt{-g} (\nabla \phi)^2~.}
In a full treatment of free field radiation from the collapsing shell  we would look for a complete set of mode solutions that on past null infinity behave as $\phi_{\omega \ell m} \sim {1\over r} Y_\ell^m(\Omega) e^{-i\omega (t+r)}$, and then compute the Bogoliubov transformation between these and analogous modes defined on the union of the event horizon and future null infinity.  This procedure cannot be carried out analytically for the case at hand.  On the other hand,  this full treatment includes lots of non-universal information pertaining to details of the collapse geometry and the grey body factors governing wave propagation in the region between the horizon and asymptotic infinity.  In order to separate these technical complications from the physics of interest here we make several standard simplifications.

We first of all reduce to the s-wave, writing (inside the shell replace $t$ by $T$)
\eq{c2}{ \phi = {1\over    r} \phit(r,t) }
so that the scalar wave equation in the regions interior  and exterior to the shell become
\eq{c3}{ {\rm interior:}\quad & (-\p_T^2 + \p_r^2) \phit =0 \cr
 {\rm exterior:}\quad &  \big(-\p_t^2 + \p_{r_*}^2 -V(r_*)\big) \phit=0 }
with $ V(r_*) = {1\over r} {d^2 r \over dr_*^2}= { (r-r_+)(r-r_-)[r_+(r-r_-)+r_-(r-r_+)]\over r^6}$.

From the definition \rf{c2}, smoothness at the origin imposes the reflecting boundary condition $\phit(r,T)|_{r=0}=0$. An infalling mode  inside the shell with dependence $e^{-i\omega (T+r)}$  simply reflects off the origin, turning into an outgoing mode proportional to $e^{-i\omega (T-r)}$.  Therefore the standard vacuum at past null infinity evolves to the vacuum with respect to the modes $e^{-i\omega (T-r)}$, which thus fixes the state emerging from the shell out into the exterior region. We are similarly in the vacuum state with respect to the infalling modes $e^{-i\omega (t+r)}$  on past null infinity to the exterior of the shell.  When we write expectation values it will always be understood that these are computed in this state.  Near the horizon this is the state that corresponds to the Unruh vacuum that is defined on the eternal black hole spacetime.

We now wish to compute scalar field correlators in this state; since the theory is free and the state is Gaussian all information is contained in the 2-point function $\langle \phi(x_1) \phi(x_2)\rangle$. Inside the shell this is the standard Minkowski space vacuum correlator (restricted to the s-wave).   In the exterior region the presence of the effective potential $V(r_*)$ in the wave equation makes the problem analytically intractable.  However, the potential goes to zero at the horizon, rises up, and then returns to zero at large $r$.  Propagation through this potential barrier gives the grey body factors, which are only analytically computable in the limit of long or short wavelengths.  Here we simply ignore the potential, with the justification being that we are interested in the physics very near the horizon where the potential does not play an important role. In any event, the effects of the potential may be incorporated later by attaching grey body factors.

Since we are in the natural vacuum state with respect to the positive frequency $(U,v)$ modes it is convenient to write the exterior metric in these coordinates.  So we convert \rf{a3c} to
\eq{c4}{ ds^2 = -f\big(r(r_*)\big){du\over dU} dUdv + r^2(r_*) d\Omega^2~.}
Ignoring the effective potential $V(r_*)$ is equivalent to using the conformally invariant  scalar field action 
\eq{c5}{ S =  2 \int\! dU dv \p_U \phit \p_v \phit~.}
The correlators then take the standard form  
\eq{c6}{ \langle \p_U \phit(U_1,v_1) \p_U \phit(U_2,v_2)\rangle  & = -{1\over 4\pi} {1\over (U_1-U_2 -i\veps)^2}  \cr
 \langle \p_v \phit(U_1,v_1) \p_v \phit(U_2,v_2) \rangle  & =  -{1\over 4\pi} {1\over (v_1-v_2 -i\veps)^2}  \cr
  \langle \p_U \phit(U_1,v_1) \p_v \phit(U_2,v_2)\rangle  & =  0~.  }
The correlator in the first line is the one of present interest.  Converting it to the $u$ coordinate in the exterior region we have
\eq{c7}{  \langle \p_u \phit(u_1,v_1) \p_u \phit(u_2,v_2)\rangle= -{1\over 4\pi}{ \left( {\p u_1 \over \p U_1}  {\p u_2 \over \p U_2}\right)^{-1} \over (U_1-U_2 -i\veps)^2}~.  }
Using \rf{b5} this can be written as an explicit function of the $U$ coordinates. However in the exterior region we are more interested in writing expressions  in terms of the $u$ coordinates. Since we can't find $U(u)$ analytically, in terms of explicit  expressions the best we can do is to focus on large $u$ (late times) and use \rf{b6} to get (corrections to this may be worked out as desired)
\eq{c8}{ \langle \p_u \phit(u_1,v_1) \p_u \phit(u_2,v_2)\rangle &\approx -{\pi T_H^2\over 4\sinh^2 \big(\pi T_H(u_1-u_2-i\veps)\big)} ~.}
This represents late time thermal behavior at temperature $T_H$; the full expression \rf{c7} contains the evolution from the standard vacuum correlator at early times to the thermal correlator at late times, corresponding to the black hole settling down to quasi-equilibrium.

Also of interest is the stress tensor expectation value
\eq{c9}{ \langle \Tt_{uu} \rangle = 2\pi \langle \p_u \phit \p_u \phit\rangle - {\rm counterterm} ~.}
The notation $\Tt_{\mu\nu}$ is used to denote the 2d stress tensor; the 4d stress tensor $T_{\mu\nu}$ includes the rescaling of the field according to \rf{c2}.
In the large $r$ asymptotic Minkowski region the counterterm defining the renormalized stress tensor corresponds to subtracting the expectation value in the $u$ vacuum.\footnote{ At finite $r$ where the metric is curved there is an additional contribution from the Weyl anomaly.}     Therefore, at large $r$ we have
\eq{c10}{ \langle \Tt_{uu}(U_1) \rangle&  = -{1\over 2} \lim_{ U_2\rt U_1} \left[  { \left( {\p u_2 \over \p U_2}  {\p u_1 \over \p U_1}\right)^{-1} \over (U_2-U_1 -i\veps)^2}  - {1 \over (u_2-u_1-i\veps)^2} \right] \cr
& = {1\over 12} \left( {\p u_1 \over \p U_1}\right)^{-2} {\rm Sch} (u_1,U_1)  }
where the Schwarzian derivative is
\eq{c11}{  {\rm Sch} (u,U)   = {u'''(U)\over u'(U)}-{3\over 2} \left( {u''(U)\over u'(U)}\right)^2~. }
This is the standard result dictated by conformal invariance, and holds for any conformal field theory upon multiplying by the central charge $c$.   The explicit result in terms of  $U$ is
\eq{c12}{  \langle \Tt_{uu}(U) \rangle& = { 1 \over 6(2r_+-U_h+U)^6}  \Big[  4r_+^2(r_+-r_-)^2 +12(r_+-r_-)^2(U_h-U) \cr
& \quad\quad  +3(3r_++r_-)(r_+-r_-)(U_h-U)^2 +2(r_++r_-)(U_h-U)^3\Big] ~,}
%
%
%
The late time  ($u\rt \infty$ or $U\rt U_h$)  result is
\eq{c14}{ \lim_{u\rt \infty}  \langle \Tt_{uu}(U) \rangle =  {(r_+-r_-)^2 \over 96 r_+^4} =  {\pi^2 T_H^2 \over 6}~,}
which represents thermal radiation at temperature $T_H$.  The energy flux smoothly rises up from zero at $u=-\infty$ to the late time thermal value \rf{c14}.  The (s-wave contribution to)  physical energy flux measured at infinity falls off as $1/r^2$ due to the rescaling \rf{c2}.

\section{Gravitationally dressed scalar field}
\label{BCMN}

Having  reviewed the standard story of free field Hawking radiation, we now turn to the incorporation of gravitational interactions. In this section we use the canonical formalism to  integrate out the gravitational field (in the s-wave) in the background of a Reissner-Nordstr\"om solution.  The approach follows that in \cite{Berger:1972pg,Unruh:1976db}, suitably generalized to include the presence of the shell.  Some useful related comments  appear in  \cite{Fischler:1990pk}.

\subsection{Action and constraints}

We start from the action for a 4d  massless real scalar field minimally coupled to gravity and electromagnetism,
\eq{d1}{S =  {1\over 4\pi}\int\! d^4x \sqrt{-g} \left[  {1\over 4G} {\cal R} -{1\over 4} F_{\mu\nu}^2  -{1\over 2} (\nabla \phi)^2 \right]~,}
  up to boundary terms.   A consistent s-wave ansatz is
  \eq{d2}{ ds^2 & = -N^2(t,r) dt^2+ L^2(t,r) (dr + N^r(t,r) dt)^2 + R^2(t,r)d\Omega^2 \cr
  A& = A_t(t,r)dt \cr
  \phi& = \phi(t,r)~,}
  where we have written the metric in ADM form with lapse $N$ and shift $N^r$.  Since  gravity and electromagnetism have no independent local degrees of freedom in the s-wave we can integrate them out to obtain an effective action for the scalar field.   To do so it is convenient to use the Hamiltonian formulation since this is designed to exhibit the constraint equations, and integrating out in this context means solving the constraints.

Following standard procedure, the action is rewritten in canonical form as
\eq{d3}{ S = \int\! dt dr \big[ \pi_\phi \dot{\phi} +\pi_R \dot{R} +\pi_L \dot{L} -N( {\cal H}^\phi_t +\Hc^G_t)-N^r ( {\cal H}^\phi_r +\Hc^G_r) \big]  -\int \! dt H_{\rm ADM}~.    }
 $(N,N^r)$ appear in the action without time derivatives and hence have no canonical momenta, instead appearing as Lagrange multiplies enforcing the constraints
 \eq{d4}{  {\cal H}^\phi_t +\Hc^G_t & = 0~,\quad  {\cal H}^\phi_r +\Hc^G_r =0~.}
Explicitly,
\eq{d3a}{ \Hc^\phi_t &= {1\over 2}\left( {\pi_\phi^2 \over LR^2}+{R^2 \over L} \phi'^2\right)~,\quad \Hc^\phi_r = \pi_\phi \phi' \cr
\Hc^G_t & = {GL\pi_L^2 \over 2R^2}-{G\pi_L \pi_R \over R} +{1\over G}\left[ \left({RR'\over L}\right)'-{R'^2 \over 2L}-{L\over 2} \right]  + { Q^2 L \over 2R^2}~,\quad \Hc^G_r =R'\pi_R -L\pi'_L~ \cr }
where $' = {d\over dr}$.
In the above we have taken the shortcut of already solving the Gauss law constraint for the electric field, yielding a Coulomb field with constant charge $Q$.  In the presence of the charged shell, which will be included later, $Q$ will be piecewise constant, vanishing inside the shell.   The ADM energy boundary term will be discussed below.

It is advantageous to define the quasilocal mass function ${\cal M}$,
\eq{d3b}{ \Mc =  {G\pi_L^2 \over 2R} +{R\over 2G} \left[ 1- \left({R'\over L}\right)^2  \right] +{Q^2 \over 2R}~,}
since then the two constraints can be rewritten as
\eq{d3c}{  \Mc' & = {R'\over L } \Hc^\phi_t   +{G\pi_L \over RL}  \Hc^\phi_r   \cr
  & R'\pi_R -L\pi'_L =- \pi_\phi \phi' ~.}

 As in related examples, the strategy  is  to  choose a gauge, solve the constraints, and then substitute back in to the action.   A convenient gauge for present purposes is\footnote{For other purposes it may be useful to use an alternative gauge choice that extends smoothly through the horizon, such as $L=1$, $R=r$. See \cite{Husain:2005gx} for a solution of the constraint equations in this gauge.}
 \eq{d5}{ \pi_L =0~,\quad R=r~.}
 The coordinate $r$ then has a clear geometric meaning in terms of the area of the $S^2$, while the vanishing of $\pi_L$ essentially corresponds to choosing Schwarzschild time.  Of course, Schwarzschild time breaks down at the black hole horizon, and so this is not a good gauge choice globally, but since our computations will take place outside the horizon this will not cause any problems.

 As a quick check, turning off the scalar field so that ${\cal M} = M$ is a constant, the constraints are easily solved to yield $\pi_R=0$ and $L^2 = \left( 1-{2GM\over r}+{GQ^2 \over r^2}\right)^{-1}$, in agreement with  the Reissner-Nordstr\"om solution \rf{a1}.  The lapse and shift are determined by the equations of motion.

 We now return to the ADM Hamiltonian $H_{\rm ADM}$.  Following Regge and Teitelboim \cite{Regge:1974zd}, its form is fixed by demanding a good variational principle.  Focusing just on the gravitational part of the bulk Hamiltonian $H_{\rm bulk}^G = \int\! dr [N \Hc_t^G + N^r \Hc^G_r]$, its variation yields the boundary term
 \eq{d6}{ \delta H_{\rm bulk}^G \big|_{\rm bndy} = -    \left[  {rN\over G L^2} \delta L \right]_{r\rt \infty}}
 where we imposed the gauge conditions \rf{d5}.    We assume large $r$ asymptotic boundary conditions that include $N \approx 1  + O(r^{-1})$ and $ L = 1 + O(r^{-1})$.   To make the action stationary we therefore need to add a boundary term whose variation cancels \rf{d6}.  Writing $L =  1+ {L_1\over r} +\ldots$  this is achieved by taking
 \eq{d7}{ H_{\rm ADM} = {L_1 \over G}~.}
 More generally, we can use that $\Mc(r)$ is the quasilocal mass, whose value at infinity gives the total energy,
 \eq{d8}{ H_{\rm ADM} = \Mc(\infty)~.}
 Indeed, evaluating \rf{d3b} at infinity subject to the asymptotic conditions assumed above gives back \rf{d7}.   For the Reissner-Nordstr\"om solution one verifies the expected result  $H_{\rm ADM}=M$.

 \subsection{Solving the constraints}

Under the gauge choice \rf{d5}  the constraints  imply the following differential equation determining $L(r)$,
\eq{d9}{ \left({r \over L^2}\right)'= 1-{GQ^2 \over r^2}- {2Gh(r)\over L^2}~,}
with
\eq{d10}{ h(r) = {1\over 2} \left( {\pi_\phi^2 \over r^2} +r^2 \phi'^2\right)~.}
The solution is
%
%
%
%
%
\eq{d11z}{ {r\over L^2(r)} & =  {r_s \over L^2(r_s) } e^{-\int_{r_s}^r {2Gh(r')\over r'} dr'}  +\int_{r_s}^r \! dr' e^{-\int_{r'}^r \! dr'' {2Gh(r'')\over r''}}- \int_{r_s}^r \! dr' {GQ^2 \over r'^2}  e^{-\int_{r'}^r \! dr'' {2Gh(r'')\over r''}}~. }
We have written the solution in terms of $L$ evaluated at the shell radius $r_s$, and the above solution strictly holds for $r>r_s$, since $Q=0$ when $r<r_s$.   In the absence of the scalar field we would have the Reissner-Nordstr\"om expression
\eq{d11zz}{ {1\over L^2(r_s)} =f(r_s)=  1-{2GM\over r_s} + {GQ^2\over r_s^2}~,}
with $f(r)$ given in \rf{a2}.
We will continue to use this expression in \rf{d11z} in the presence of the scalar field.  Although this amounts to ignoring the gravitational effects of the scalar field inside the shell we are really only interested in the scalar field interactions in the region outside the shell (and more specifically in the near-horizon region) since this is where the near-extremal enhancement arises,  so this step will not affect our results.  So going forward, for $r>r_s$ we take  the solution to be \rf{d11z} with \rf{d11zz}.

After some integration by parts (see appendix \rff{parts}) we can write 
%
%
%
\eq{d11v}{ {1\over L^2(r)} & = f(r) -{2G\over r} \int_{r_s}^r \! dr' f(r') h(r')  e^{-\int_{r'}^r \! dr'' {2Gh(r'')\over r''}}~.  }
Using this expression the quasilocal mass function $\Mc(r)$ takes the form 
%
%
%
\eq{d20}{ \Mc(r)
& =    M +  \int_{r_s}^r \! dr' f(r') h(r')  e^{-\int_{r'}^r \! dr'' {2Gh(r'')\over r''}} ~,   }
whose large $r$ limit then gives  the ADM Hamiltonian as
%
%
%
\eq{d21}{ H_{\rm ADM} & =   M + \int_{r_s}^\infty \! dr' f(r') h(r')  e^{-\int_{r'}^\infty \! dr'' {2Gh(r'')\over r''}}~.  }
This is the main result of this section: a Hamiltonian which captures all the s-wave gravitational interactions of the scalar field with itself and with the background geometry. It is of course nonlocal in space on account of the long range nature of gravity, but when inserted back into \rf{d3} it yields an action local in time, as follows from the absence of dynamical degrees of freedom in the pure gravity sector.

\subsection{Effective scalar field action}

Under our gauge choice $\pi_L=0$ and $R=r$, after solving the constraints  the action takes the first order form
\eq{q1}{ S = \int\! dt dr  \pi_\phi \dot{\phi}  -\int\! dt H_{\rm ADM} }
with $H_{\rm ADM}$ given in \rf{d21}.  To obtain the second order form we should solve for $\pi_\phi$ and substitute back in or, more properly, integrate over $\pi_\phi$ in the path integral.  Neither of these steps can be carried out analytically so we  employ perturbation theory, working order by order in the function $h(r)$ which controls the gravitational response.

As a first check consider the expansion to first order in $h$,
\eq{d22}{ H_{\rm ADM} & = M+  \int_{r_s}^\infty      \! dr'  f(r') h(r') +O(h^2) }
where $f(r') = 1-{2GM\over r'}+{GQ^2 \over r'^2}$  is equal to $-g_{tt}$ in the Reissner-Nordstr\"om metric.  This represents the free field energy of the scalar field multiplied by the redshift factor.
So at first order the action is
 \eq{d24}{ S^{(1)} & = \int\! dt dr \pi_\phi \dot{\phi} -{1\over 2} \int\! dt \int_{r_s}^\infty\! dr' f(r')\left( {\pi_\phi^2 \over r^2} +r^2 \phi'^2\right)~. }
Solving for $\pi_\phi$ gives
\eq{d25a}{ \pi_\phi(r,t)  = {r^2 \over f(r) }\dot{\phi}(r,t) + O(\phi^3)~.}
Substituting back in gives the action quadratic in fields,
\eq{d26}{ S^{(1)} &=  {1\over 2} \int\! dt \int_{r_s}^\infty \! dr r^2 \left( {\dot{\phi}^2 \over f}- f\phi'^2\right)}
 which is the expected free field s-wave action on the Reissner-Nordstr\"om background, $S^{(1)} =-{1\over 8\pi } \int\! d^4x  g^{\mu\nu} \p_\mu \phi \p_\nu \phi$.

 Expanding the Hamiltonian  to order $h^2$ gives 
\eq{d23}{ H_{\rm ADM}^{(2)} & =-2G\int_{r_s}^\infty\! dr f(r) h(r) \int_{r}^\infty\! dr' {h(r')\over r'} ~.}
%
%
%
%
%
If we only want the action to quartic order in the scalar field we can just plug the solution \rf{d25a} into the quartic Hamiltonian \rf{d23}.   To this end, note that \rf{d10} yields
\eq{d27x}{ h(r) =  {1\over 2} \left( {r^2 \over f^2} \dot{\phi}^2 + r^2 \phi'^2 \right)}
so we simply use this relation in \rf{d23} to obtain the quartic interaction terms.

To summarize, the action to quartic order is 
%
%
%
\eq{d29}{ S &= \int\! dt \int_{r_s}^\infty\! dr \Bigg[ {1\over 2}r^2 \left( {1\over f}\dot{\phi}^2-f\phi'^2 \right)  -2G f(r) h(r) \int_{r}^\infty\! dr' {h(r')\over r'}\Bigg]~.}
with $h(r)$ given by \rf{d27x}.

\subsection{Scalar action in the near-extremal, near-horizon limit}

We now focus attention on the near-extremal limit with interactions taking place near the horizon.  To this end we implement the redefinitions  \rf{a4}-\rf{a6}, reproduced here for convenience
\eq{d29x}{ r_\pm &= r_0 \pm \lambda \rho_h  \cr
r& = r_0 +\lambda \rho \cr
t& = {r_0^2 \over \lambda }\tau~. }
We similarly rescale the light cone coordinates, $(u=t-r_*,v=t+r_*)$ with $r_*$ given by \rf{a4x}, as  
\eq{d3g}{ u = {r_0^2 \over \lambda } \ut~,\quad v = {r_0^2 \over \lambda } \vt }
which as $\lambda \rt 0$ implies
\eq{d31ab}{\ut  = \tau -{1\over 2\rho_h} \ln\left({\rho-\rho_h\over \rho+\rho_h}\right)~,\quad \vt  = \tau +{1\over 2\rho_h} \ln\left({\rho-\rho_h\over \rho+\rho_h}\right)~. }
Finally, we rescale   the scalar field as (see \rf{c2})
\eq{d31a}{ \phi = {1\over r_0} \phit~. }
After inserting all these rescalings into the action \rf{d29} we take $\lambda\rt0$, keeping the leading forms of the quadratic and quartic terms.  This results in the action
\eq{d31}{  S& = \int\! d\tau \Bigg[ 2 \int_{\rho_s}^\infty {d\rho\over \rho^2-\rho_h^2} \p_{\ut} \phit \p_{\vt} \phit \cr
& \quad +{2G\over   r_0\lambda} \int_{\rho_s}^\infty {d\rho \over \rho^2- \rho_h^2}  \int_{\rho}^\infty {d\rho' \over \rho'^2-\rho_h^2} { \big[ (\p_{\ut} \phit)^2+(\p_{\vt} \phit)^2 \big]\big[ (\p_{\ut'} \phit)^2+(\p_{\vt'} \phit)^2 \big]\over \rho'^2-\rho_h^2   }  \Bigg]~,}
where $\p_{\ut} \phit = \p_{\ut} \phit(\ut,\vt)$, $\p_{\ut'} \phit = \p_{\ut'} \phit(\ut',\vt')$, etc.   Importantly, the terms in the second line are proportional to ${G\over \lambda} \sim {G\over T_H}$, thus exhibiting a low temperature enhancement.  

The action \rf{d31} controls gravitational interactions of the scalar field in the near-horizon region, to first order in $G$.

\section{Gravitational perturbation theory}
\label{pert}

We now turn to the computation of    gravitational corrections to scalar field correlation functions to first order in $G$, focusing on  corrections that behave as ${G\over T_H}$ in the low temperature regime.  To probe the outgoing Hawking  radiation it will suffice to compute correlation functions  involving only $\p_{\ut} \phit$.
%
%
Ultimately we are interested in the radiation that escapes to the asymptotically flat region.  As discussed above, between the near-horizon region and asymptotic infinity we have (to good approximation) free fields propagating in the presence of the effective potential.    What we actually compute here are correlation functions at the boundary of the near-horizon region, which corresponds to $\rho \rt \infty$ after the small $\lambda$ limit has been taken. The free field equations, with grey body factors computed from the effective potential, may be used to connect these near-horizon results to observables in the asymptotically flat region, but this is not considered here.   Such matching procedures are standard in black hole computations.

\subsection{Perturbative setup}

We consider  correlators
\eq{e1}{ G_n(\ut_1,\vt_1; \ldots ; \ut_n ,\vt_n)=  \langle  \p_{\ut_1} \phit_H(\ut_1, \vt_1) \ldots  \p_{\ut_n} \phit_H(\ut_n, \vt_n)\rangle}
 in the Heisenberg picture, with the state taken to be  $|Uv\rangle$ as defined above.   The operators will be taken to lie on some large $\rho$ surface of the near-horizon region, and to get fully analytical results we will eventually take the times $\tau_i$ to be large as well.

To proceed, we pass to the interaction picture, writing  free fields as $\phit$, related to the Heisenberg fields $\phi_H$ as (suppressing the spatial dependence)
\eq{e2}{ \phit_H(t) = U^\dagger(t,t_0)\phit(t) U(t,t_0)}
where
\eq{e3}{ U(t,t_0)= T\left[ e^{-i\int_{t_0}^t\! dt' H_I(t')}\right]~.}
The time $t_0$ will be taken to lie in the far past where the incoming state is defined.  We will work to first order in the interaction, so
\eq{e4}{  \phi_H(t) = \phit(t) +i \int_{t_0}^t\! dt'  [H_I(t'),\phit(t)]~.}
We  insert this in \rf{e1} and then Wick contract using the free field correlators.

To evaluate \rf{e4} we use the basic free field commutators
\eq{e5}{  [\p_{\ut} \phit(\ut), \p_{\ut'} \phit(\ut')]&= {i\over 2} \p_{\ut}  \delta(\ut-\ut') \cr
 [\p_{\ut} \phit(\ut), \p_{\vt'} \phit(\vt')]&=0~. }
Our interaction Hamiltonian is
\eq{e6}{ H_I & = - {2G\over   r_0\lambda} \int_{\rho_s}^\infty {d\rho \over \rho^2- \rho_h^2}  \int_{\rho}^\infty {d\rho' \over \rho'^2-\rho_h^2} { \big[ (\p_{\ut} \phit)^2+(\p_{\vt} \phit)^2 \big]\big[ (\p_{\ut'} \phit)^2+(\p_{\vt'} \phit)^2 \big]\over \rho'^2-\rho_h^2   }  \Bigg]~.}
Writing the explicit terms that appear in the commutator with $\p_{\ut_1} \phit(\ut_1)$ we have
\eq{e6a}{& i\int_{\tau_0}^{\tau_1}\! d\tau''[H_I(\tau'') ,  \p_{\ut_1} \phit(\ut_1) ] \cr
& = -{2G\over r_0 \lambda}  \int_{\tau_0}^{\tau_1}\! d\tau'' \int_{\rho_{s,1}}^\infty {d\rho \over \rho^2- \rho_h^2}  \int_{\rho}^\infty {d\rho' \over \rho'^2-\rho_h^2}   { [i(\p_{\ut} \phit(\ut))^2,\p_{\ut_1} \phit(\ut_1) ] \big( (\p_{\ut'} \phit(\ut'))^2 + (\p_{\vt'} \phit(\vt'))^2 \big)  \over \rho'^2-\rho_h^2 } \cr
& \quad -{2G\over r_0 \lambda}  \int_{\tau_0}^{\tau_1}\! d\tau'' \int_{\rho_s(\tau_1)}^\infty {d\rho \over \rho^2- \rho_h^2}  \int_{\rho}^\infty {d\rho' \over \rho'^2-\rho_h^2}   { \big( (\p_{\ut} \phit(\ut))^2 + (\p_{\vt} \phit(\vt))^2 \big) [i(\p_{\ut'} \phit(\ut'))^2,\p_{\ut_1} \phit(\ut_1) ] \over \rho'^2-\rho_h^2 }~. \cr
 }
From \rf{e5} we have
\eq{e6b}{ [i(\p_{\ut} \phit(\ut))^2,\p_{\ut_1} \phit(\ut_1) ] &=-\p_{\ut} \phit(\ut) \p_{\ut} \delta(\ut-\ut_1)\cr
 [i(\p_{\ut'} \phit(\ut'))^2,\p_{\ut_1} \phit(\ut_1) ] &=-\p_{\ut'} \phit(\ut') \p_{\ut'} \delta(\ut'-\ut_1)~.  }
Using these delta functions and the fact that $\rho'\geq \rho$ we can isolate the region of integration that survives in the two lines of \rf{e6a}.   These regions are shown in figure \rff{regions}.
\begin{figure}[H]
\centering
\includegraphics[scale=0.35]{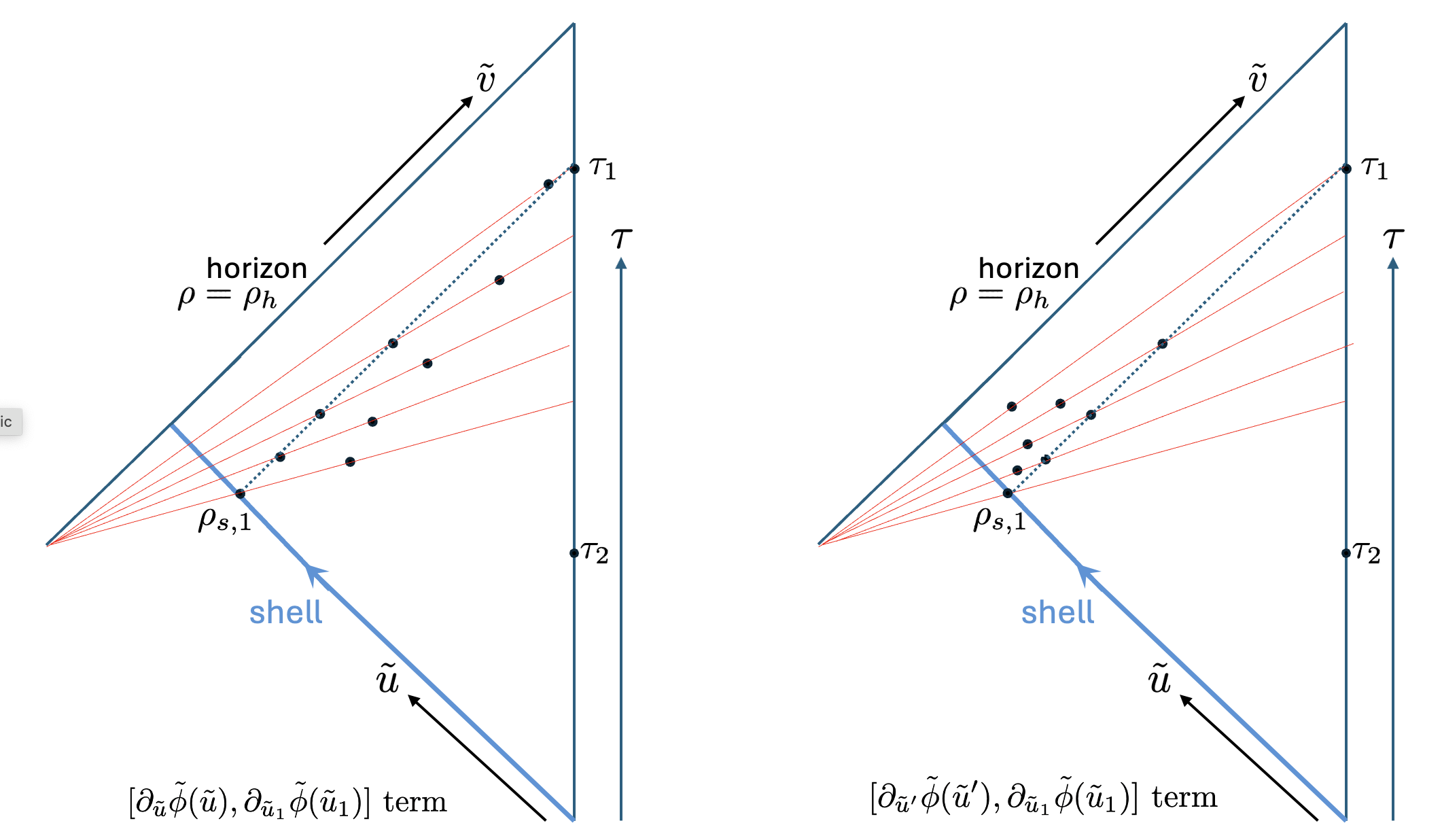}
\caption{  Integration regions contributing in the second and third lines of \rf{e6a}.   Lines of constant $\tau$ are in red.  The dotted line is a null ray of constant $\ut=\ut_1$.   The black dots denote typical values of $\rho$ and $\rho'$ that contribute to the integrals, where $\rho \leq \rho'$.    On the left panel  we have a commutator involving  $\p_{\ut} \phit(\ut)$ with  $\p_{\ut_1} \phit(\ut_1)$,  which forces $\ut=\ut_1$.  Similarly, in the right panel it is $\p_{\ut'} \phit(\ut')$ that  appears in a commutator with  $\p_{\ut_1} \phit(\ut_1)$, forcing $\ut' = \ut_1$.   The lower bound on the $\rho$ integral is set by the shell location.      }
\label{regions}
\end{figure}
As shown in the figure, a finite range of $\tau''$ contributes, with limits set by the operator insertion time $\tau_1$ and the time at which  that boundary point's past light cone intersects with the shell trajectory.

As indicated,  $\rho_{s,1}$ is the radial location at which the shell trajectory crosses the null line $\ut=\ut_1$.   To work out its explicit value we use that in the AdS$_2$ metric $ds^2 = - (\rho^2-\rho_h^2) d\tau^2 +{d\rho^2 \over \rho^2 -\rho_h^2}$ a past directed null geodesic emanating from the boundary point $(\tau_1, \rho_b)$, with $\rho_b\gg \rho_h$, obeys
\eq{e9}{
\tau \approx \tau_1 +{1\over 2\rho_h}\ln\left( {\rho-\rho_h \over \rho+\rho_h}\right) ~.}
On the other hand consider the trajectory of the infalling null shell.  This obeys (for a particular choice of infall time, which is immaterial due to time translation invariance)
\eq{e10}{ \tau_{\rm shell} \approx  -{1\over 2\rho_h} \ln\left( {\rho-\rho_h \over \rho+\rho_h}\right) ~.}
These two null trajectories cross at
\eq{e11}{ \rho_{s,1} & = \rho_h \coth (\rho_h \tau_1/2) \approx \rho_h +2 \rho_h e^{- \rho_h \tau_1}}
where in the second equality we expanded for large $\tau_1$.  \rf{e11} gives the lower $\rho$ cutoff to be used in \rf{e6a}. It approaches the horizon at late boundary time.

In figure \rff{regions} it is apparent that the lower limit for $\rho $ takes the value $\rho_{s,1}$ in the left panel, and $\rho_s(\tau_1)$ in the right panel, where $\rho_s(\tau)$ denotes the shell trajectory.

Taking the above discussion into account, we can now use the delta functions to perform the $\tau''$ integration and obtain
\eq{e11a}{ &i\int_{\tau_0}^{\tau_1}\! d\tau''[H_I(\tau'') ,  \p_{u_1} \phit(u_1) ]  \cr
& =  -{G \over r_0\lambda}\p^2_{\ut_1} \phit(\ut_1) \Bigg( \int_{\rho_{s,1}}^\infty {d\rho \over \rho^2- \rho_h^2}  \int_{\rho}^\infty {d\rho' \over \rho'^2-\rho_h^2}  {\big( (\p_{\ut'} \phit(\ut'))^2 + (\p_{\vt'} \phit(\vt'))^2 \big)  \over \rho'^2-\rho_h^2  } \cr
& \quad\quad\quad\quad\quad\quad\quad\quad  + \int_{\rho_s(\tau_1) }^\infty {d\rho \over \rho^2- \rho_h^2}  \int_{\rho}^\infty {d\rho' \over \rho'^2-\rho_h^2} {\big( (\p_{\ut} \phit(\ut))^2 + (\p_{\vt} \phit(\vt))^2 \big)  \over \rho'^2-\rho_h^2}  \Bigg)~.        }
Strictly speaking $\p^2_{\ut_1} \phit(\ut_1)$ should lie to the right of the other operators in the second line, but this won't matter for present purposes.  Also, the precise values of $(\ut,\vt,\ut',\vt')$  implied by the delta function will not be needed, but for completeness the explicit expressions are
\eq{e8}{ \ut &  = \tau_1 - {1\over 2\rho_h} \ln\left( {\rho-\rho_h\over \rho+\rho_h}\right) +{1\over 2\rho_h} \ln\left( {\rho'-\rho_h\over \rho'+\rho_h}\right) \cr
\ut' &  = \tau_1 + {1\over 2\rho_h} \ln\left( {\rho-\rho_h\over \rho+\rho_h}\right)-{1\over 2\rho_h} \ln\left( {\rho'-\rho_h\over \rho'+\rho_h}\right) \cr
\vt& =\vt' =  \tau_1 + {1\over 2\rho_h} \ln\left( {\rho-\rho_h\over \rho+\rho_h}\right)+{1\over 2\rho_h} \ln\left( {\rho'-\rho_h\over \rho'+\rho_h}\right)~. }
Here we have taken $\ut_1$ to lie at  large $\rho$ boundary  where $\ut_1 \approx  \tau_1$.

\subsection{Perturbative evaluation of late time correlators }

For use in carrying out perturbation theory in the near-horizon geometry  we now collect the free field building blocks.  The basic bulk 2-point function is \rf{c7}, however in order to obtain analytical results we restrict here to the late time regime where we have \rf{c8}, which in the rescaled near-horizon variables is
\eq{e12}{ G_2^{(tree)}(\ut_1,\ut_2)=  \langle \p_{\ut} \phit(\ut_1) \p_{\ut} \phit(\ut_2)\rangle &\approx -{\rho_h^2 \over 16\pi \sinh^2 \left({\rho_h (\ut_1-\ut_2-i\veps)\over 2} \right)  }~.}
If one of the points lies at the large $\rho$ boundary then we can replace $\ut$ by $\tau$.

We will also make use of self-contractions in the bulk.  As discussed in Appendix \rff{stress} in the near-horizon region we have
\eq{e13}{ \langle \Tt_{\ut\ut}\rangle =0~,\quad  \langle \Tt_{\vt\vt}\rangle =-{\rho_h^2 \over 24 }~.}
which implies the self-contractions\footnote{This involves the assumptions that we should renormalize the operators $(\p_{\ut} \phit)^2$ and $(\p_{\vt} \phit)^2$ that appear in the interaction Hamiltonian by viewing them as two components of the stress tensor $\Tt_{\mu\nu}$.  Doing so means we can use properties of the stress tensor, namely its conservation and tensor property, to fix renormalization ambiguities.  Ultimately, this requires justification, for instance by imposing general covariance.}
\eq{e14}{ \langle \p_{\ut} \phit(\ut) \p_{\ut}\phit(\ut)\rangle =0~,\quad  \langle \p_{\vt} \phit(\vt) \p_{\vt}\phit(\vt)\rangle =-{\rho_h^2 \over 48 \pi  }~.}
We also recall that in the near-horizon region
\eq{e15}{ \ut  = \tau -{1\over 2\rho_h} \ln\left({\rho-\rho_h\over \rho+\rho_h}\right)~.}

\subsubsection{One-loop correction to 2-point function}

Using \rf{e11a}, the one-loop contribution to the 2-point function $G_2 = \langle \p_{\ut_1} \phit_H(\ut_1)\p_{\ut_2} \phit_H(\ut_2)\rangle $ is
\eq{e16}{ &G_2^{({\rm one-loop})}  (\ut_1,\ut_2)\cr
&=   -{G \over r_0\lambda} \Bigg( \int_{\rho_{s,1}}^\infty {d\rho \over \rho^2- \rho_h^2}  \int_{\rho}^\infty {d\rho' \over (\rho'^2-\rho_h^2)^2} \big\langle  \p^2_{\ut_1} \phit(\ut_1) \big[ (\p_{\ut'} \phit(\ut'))^2 + (\p_{\vt'} \phit(\vt'))^2 \big]\p_{\ut_2} \phit(\ut_2) \big\rangle \cr
& \quad  \quad  \quad + \int_{\rho_s(\tau_1)}^\infty {d\rho \over \rho^2- \rho_h^2}  \int_{\rho}^\infty {d\rho' \over (\rho'^2-\rho_h^2)^2} \big\langle  \p^2_{\ut_1} \phit(\ut_1) \big[ (\p_{\ut} \phit(\ut))^2 + (\p_{\vt} \phit(\vt))^2 \big]\p_{\ut_2} \phit(\ut_2) \big\rangle \Bigg) \cr
 &  \quad \quad \quad+  [\big(\rho_{s,1},\rho_s(\tau_1)\big) ~\rt \big(\rho_{s,2},\rho_s(\tau_2)\big)]~.        }
There are two distinct types of Wick contractions, corresponding to the following two distinct one-loop diagrams in the underlying theory of gravity coupled to the scalar field,
\eq{q1z}{
\begin{tikzpicture}[baseline=0.1cm,scale=0.25]
\begin{feynhand}
\vertex (a) at (0,0); \vertex (b) at (10,8);  \vertex (c) at (10,-8);
\propag [plain]  (a) to (b);
\propag [plain]  (a) to (c);
\propag [plain]  (b) to (c);
\vertex (d) at (4,0); \vertex (e) at (7,0);
\propag [photon]  (d) to (e);
\vertex (f) at (10,3)   ;\vertex (g) at (10,-3);
\vertex (f1) at (10.9,3)   {$\tau_1$} ;\vertex (g1) at (10.9,-3)   {$\tau_2$};
\propag [plain]  (e) to (f); \propag [plain]  (e) to (g);
\draw (3,0) circle (1cm);
\vertex (a5) at (15,0); \vertex (b5) at (25,8);  \vertex (c5) at (25,-8);
\propag [plain]  (a5) to (b5);
\propag [plain]  (a5) to (c5);
\propag [plain]  (b5) to (c5);
\vertex (f5) at (25,4)   ;\vertex (g5) at (25,-4);
\vertex (f15) at (25.9,4)   {$\tau_1$} ;\vertex (g15) at (25.9,-4)   {$\tau_2$};\propag[plain] (f5)  to [half right, looseness =2.1]   (g5);
\vertex (m5) at (22.2,3.34);\vertex (n5) at (22.2,-3.34);
\propag [photon]  (m5) to (n5);
\vertex (x)   at (5,-8)   {(a)};\vertex (x)   at (20,-8)   {(b)};
\end{feynhand}
\end{tikzpicture}
}
We focus here on diagram (a), which is simpler to evaluate. In  diagram (a) we  Wick contract $\phit(\ut_1)$ and $\phit(\ut_2)$, and self-contract the integrated fields. The self-contractions are given by \rf{e14} and are in particular constant and so can be brought out of the integral.  We then arrive at
\eq{e17}{ G_{2,a}^{({\rm one-loop})}& ={G\rho_h^2 \over 48\pi r_0 \lambda} \Big[ I(\tau_1) \p_{\ut_1} G_{2}^{({\rm  tree})}(\ut_1,\ut_2) +I(\tau_2) \p_{\ut_2} G_{2}^{({\rm  tree})}(\ut_1,\ut_2) \Big]    }
with
\eq{e18}{ I(\tau_k) &=
  \int_{\rho_{s,k}}^\infty {d\rho \over \rho^2- \rho_h^2}  \int_{\rho}^\infty {d\rho' \over (\rho'^2-\rho_h^2)^2}+ \int_{\rho_s(\tau_k)}^\infty {d\rho \over \rho^2- \rho_h^2}  \int_{\rho}^\infty {d\rho' \over (\rho'^2-\rho_h^2)^2} \cr
  & =  {1\over 4 \rho_h^2}{1\over \rho_{s,k}^2 -\rho_h^2} - \left[ {1\over 4 \rho_h^2}\ln \left( {\rho_{s,k} -\rho_h \over \rho_{s,k} + \rho_h}\right)\right]^2  +  [\rho_{s,k} \rt \rho_s(\tau_k) ]~.}
The integrals grow at late times since $\rho_{s,k}$ and $\rho_s(\tau_k)$ approach $\rho_h$.    In particular, using (see \rf{e11})  the late time behavior $\rho_{s,k} \approx \rho_h +2 \rho_h e^{- \rho_h \tau_k}$, we have
\eq{e18a}{   {1\over 4 \rho_h^2}{1\over \rho_{s,k}^2 -\rho_h^2} - \left[ {1\over 4 \rho_h^2}\ln \left( {\rho_{s,k} -\rho_h \over \rho_{s,k} + \rho_h}\right)\right]^2
  \approx  {1\over 16\rho_h^4} e^{\rho_h \tau_k} -{1\over 16 \rho_h^4} ( \tau_k)^2~. }
The $\rho_s(\tau_k)$ contribution is messier but has the same structure consisting of a part that grows as $e^{\rho_h \tau_k}$ and a part that grows quadratically in $\tau_k$, along with subleading terms.  Without keeping track of the numerical coefficients we thus write
\eq{e18b}{ I(\tau_k)  \approx {C_1 \over \rho_h^4}e^{\rho_h \tau_k} + {C_2 \over \rho_h^4}  ( \tau_k)^2~. }

Let's consider the regime  $\rho_h \tau_1 \gg \rho_h \tau_2 \gg 1$.  Then  the first term in  \rf{e17} dominates and the  tree level correlator \rf{e12} is well approximated by the quasi-normal decay,
\eq{e20}{ G_2^{\rm (tree)}(\ut_1,\ut_2) \approx - {\rho_h^2 \over 4\pi } e^{-\rho_h(\ut_1 -\ut_2)}~,
}
so that (up to terms subleading at large $\tau_1$)
\eq{e21}{ { G_{2,a}^{({\rm one-loop})} \over  G_{2,a}^{({\rm tree})} } \approx -{C_1\over 48 \pi^2} {G\over T_H}e^{\rho_h \tau_1}  +{C_2\over 48 \pi^2}{G\over r_0^3 T_H} \rho_H^2 \tau_1^2~.    }
We discuss this result in the next section. Here we note that it's straightforward to write down an expression for diagram (b) in \rf{q1z}, but we have not performed the integrals;  similarly for the 4-point function.

\subsection{Growing terms in the 2-point function at one-loop}

Diagram (a) in  \rf{q1z} contains terms \rf{e21} that grow at large $\tau$.  We now discuss the interpretation of these terms, starting with the quadratically growing term, which  can be written  in terms of the original Schwarzschild time $t$ as
\eq{e22}{ { G_{2,a}^{({\rm one-loop})} \over  G_{2,a}^{({\rm tree})} }\Big|_{\rm quadratic}  \approx {C_2\over 3\pi^2 }{ G\over r_0^3 T_H } (2\pi T_H t_1)^2~.}
This has the same form as the term which dominates the one-loop Schwarzian correlator at late time \cite{Maldacena:2016upp}.  In that reference the presence of this growing term was given a simple interpretation in terms of temperature (or energy) fluctuations of the black hole.      For $t_1$ of order $1/T_H$ the one-loop correction becomes of order the tree level term at temperature $T_H \sim {G\over r_0^3}$, which agrees with the  temperature inferred from the breakdown of  thermodynamics, and similarly appears in the Schwarzian description.

Next we consider the exponentially growing term in \rf{e21}.  No such term is present in the one-loop Schwarzian computation of \cite{Maldacena:2016upp}.  The origin of this term in our case comes from the integration region in which the vertices are approaching the horizon, a region which would also be present in \cite{Maldacena:2016upp}.   However, a key difference is that the Euclidean computation in \cite{Maldacena:2016upp} corresponds to a state of thermal equilibrium, which from the asymptotically flat black hole spacetime point of view means that we are feeding energy into the black hole at the same rate as it radiates.   On the other hand, in our setup we assumed the vacuum state in the far past.  At the same time,  the value of diagram (a) is proportional to $\langle T_{\vt\vt}\rangle$  evaluated in the near-horizon region, as given by \rf{z6}.
$\langle T_{\vt\vt}\rangle$  represents the negative energy flowing into the horizon causing the black hole to shrink, and is precisely the contribution that one would cancel to maintain equilibrium, by sending in radiation from past null infinity at the Hawking temperature.    So we expect that if we were to include such an incoming state of thermal radiation then this diagram would vanish at late times, leaving contribution (b).   On the other hand, if we stick with the state of no incoming radiation, as is perfectly reasonable physically, then in order to ``resum" the exponentially growing term we should presumably look for a self-consistent background solution representing the shrinking black hole, and for which the tadpole diagram is suppressed.  Finally, we expect that the diagram (b) is insensitive to this issue, as is the 4-point function, as they do not involve tadpoles.

\section{Discussion }
\label{discussion}

The main result of this work is a Hamiltonian framework that incorporates large near-horizon gravitational fluctuations into the analysis of Hawking radiation and scattering.  The first steps in the evaluation of correlators were taken, and it will be interesting to work out the full 2-point and 4-point correlation functions and extract their physical implications.  At the classical level the full effective Hamiltonian was written explicitly, but working out its implications at the quantum level presents a novel challenge in effective field theory in order to isolate the universal IR effects that it encodes.  We plan to return to this problem.

It would be very useful to understand the precise relation to the Schwarzian action describing JT gravity, and more specifically its use in  computing gravitationally dressed scalar correlators. The Schwarzian  formulation is very efficient since it reduces the problem to the boundary; importing that feature to the Lorentzian context developed here would be a technical advantage.

Finally, we comment on some previous work regarding the effect of  gravitational interactions on Hawking radiation, not geared specifically to the context of near-extremal black holes.   In \cite{Stephens:1993an} and related papers it was argued that strong gravitational interactions near the horizon invalidate the semiclassical approximation.  Interactions between infalling matter and outgoing radiation were studied in \cite{Kiem:1995iy}.  More recently, these effects were related to chaotic properties of black holes, e.g. \cite{Shenker:2013pqa,Polchinski:2015cea}.   These papers involve shock wave interactions between particles of large relative boost, and it would be interesting to understand how this emerges from our all-orders Hamiltonian.    The papers \cite{Kraus:1994by,Kraus:1994fj,Keski-Vakkuri:1996wom} were aimed at understanding black hole emission of highly energetic quanta, those whose gravitational self-interaction has a large effect.  By thinking of such particles as self-gravitating shells,  it was found that thermal Boltzmann factors $e^{-\beta \omega}$ were replaced by $e^{-\Delta S}$, where $\Delta S$ is the change in black hole entropy between a black hole of mass $M$ and one of mass $M-\omega$, an expression which reduces to the Boltzmann factor for small $\omega$.   However, the precise meaning and  regime of validity of this result is unclear, and so it would be useful to recast it in terms of correlators.  The approach developed in this work may be suitable for that purpose.

\section*{Acknowledgments}

Work  supported in part by the National Science Foundation grant PHY-2209700.   I thank Richard Myers for discussions, and   the Centro de Ciencias de Benasque Pedro Pascual and the organizers of the ``Gravity - New quantum and string perspectives” workshop for their hospitality during the course of this work.

\appendix

\section{Stress tensor expectation value}
\label{stress}

Here we review some basic formulas for the quantum stress tensor  on a 2d metric.   See \cite{Birrell:1982ix}   or, e.g., \cite{Balbinot:2023vcm} for a discussion in the Reissner-Nordstr\"om context.

Consider a conformal field theory of central charge $c$ on the metric
\eq{z1}{ ds^2=-f(r)(dt^2-dr_*^2) = -f(r) dudv}
with ${dr_* \over dr}= {1\over f(r)}$, $u=t-r_*$ and $v=t+r_*$.   The expectation value of the stress tensor is determined by the Weyl anomaly and a choice of quantum state.   Here we consider the vacuum state defined with respect to the coordinates $U(u)$ and $v$, meaning that we use these coordinates to distinguish creation and annihilation operators.  We write this state as $|Uv\rangle$.   In this state we have
\eq{z2}{  \langle Uv| \Tt_{uu}|Uv\rangle & =- {c\over 6}\left({1\over 16} f'(r)^2 -{1\over 8} f(r)f''(r)   \right)  -{c\over 12}{\rm Sch}(U,u) \cr
 \langle Uv| \Tt_{vv}|Uv\rangle   & =  - {c\over 6}\left({1\over 16} f'(r)^2 -{1\over 8} f(r)f''(r)   \right)\cr
 \langle Uv| \Tt_{uv}|Uv\rangle   & = -{c\over 6} \left( -{1\over 8} f(r)f''(r) \right)}
with
\eq{z3}{ {\rm Sch}(U,u) = { U''' \over U'}-{3\over 2} \left( {U'' \over U'}\right)^2~,}
and where, as in the main text, the tildes are there to distinguish the 2d stress tensor from the 4d stress tensor.
The contributions involving $f$ can be thought of as coming from the Weyl anomaly; they would be present also in the $|uv\rangle$ vacuum state.  The Schwarzian contribution comes from using the $U$ vacuum state.

In the Reissner-Nordstr\"om case we have $f(r) = {(r-r_+)(r-r_-)\over r^2}$.  The resulting expressions for the stress tensor are somewhat messy and will not be written here, but we emphasize a few key points.

First consider $\Tt_{uu}$ and use our late time expression  \rf{b6},  $U\approx U_h-C e^{-2\pi T_U u}$.   In the near-horizon and asymptotic limits the formulas above then give
\eq{z4}{  \langle Uv| \Tt_{uu}|Uv\rangle & = \left\{  \begin{array}{cc}
                                                 O\big( (r-r_+)^2 \big)  &{\rm as~}  r\rt r_+ \\
                                                 {c\pi^2  T_H^2\over 6} &{\rm as~}  r\rt \infty
                                               \end{array}
\right.   }
The vanishing behavior as the horizon is approached is needed in order to obtain a finite result for the stress tensor expressed in terms of the $U$ coordinate which, unlike $u$, is well behaved at the horizon.   The large $r$ result, which comes entirely from the Schwarzian term,  gives back our previous expression \rf{c14}.

Next consider $\Tt_{vv}$. It is apparent from \rf{z2} that
\eq{z5}{ \lim_{r\rt r_+ } \langle Uv| \Tt_{vv}|Uv\rangle  &= -{c\over 96} [f'(r_+)]^2  = - {c\pi^2  T_H^2\over 6}~. }
This represents the flux of negative energy flowing into the horizon, causing the black hole to shrink.  The $v$ coordinate is well defined at the horizon, so this represents a nonsingular stress tensor.

We also write the result in the near-extremal limit, $\lambda \rt 0$.  Using $\vt ={\lambda\over r_0^2} v$ from  \rf{d31} and that $T_H \approx {\rho_h \lambda \over 2\pi r_0^2}$ as $\lambda \rt 0$ we have
\eq{z6}{  \langle Uv| \Tt_{\vt\vt}|Uv\rangle \approx  -{c\rho_h^2 \over 24 }~. }
On the other hand $ \langle Uv| \Tt_{\ut\ut}|Uv\rangle = O(\lambda)$ and so will be taken to vanish. This result is understood as in the comment below \rf{z4}.

\section{Rewriting of gravitational Hamiltonian }
\label{parts}

Starting from \rf{d11z} we integrate by parts as
\eq{d11y}{ \int_{r_s}^r \! dr' e^{-\int_{r'}^r \! dr'' {2Gh(r'')\over r''}} & = \int_{r_s}^r \! dr' {d\over dr'}(r') e^{-\int_{r'}^r \! dr'' {2Gh(r'')\over r''}}  \cr
& = -\int_{r_s}^r\! dr' 2Gh(r') e^{-\int_{r'}^r \! dr'' {2Gh(r'')\over r''}}   +r-r_s e^{-\int_{r_s}^r \! dr' {2Gh(r')\over r'}}   }
and
\eq{d11x}{ &- \int_{r_s}^r \! dr' {GQ^2 \over r'^2}  e^{-\int_{r'}^r \! dr'' {2Gh(r'')\over r''}} =  \int_{r_s}^r \! dr'  {d\over dr'} \left( {GQ^2 \over r'}\right)   e^{-\int_{r'}^r \! dr'' {2Gh(r'')\over r''}} \cr
& \quad  = - \int_{r_s}^r \! dr' {GQ^2 \over r'^2}  2Gh(r') e^{-\int_{r'}^r \! dr'' {2Gh(r'')\over r''}} +{GQ^2 \over r} -{GQ^2 \over r_s} e^{-\int_{r_s}^r \! dr' {2Gh(r')\over r'}}~.  }
We then write
\eq{dllw}{ {r_s\over L^2(r_s)} = r_s -2GM+{GQ^2\over r_s}}
to get
\eq{d11vz}{ {r\over L^2(r)} & = r +{GQ^2 \over r} -2GM e^{-\int_{r_s}^r \! dr' {2Gh(r')\over r'}}  -\int_{r_s}^r \! dr' \left( 1+{GQ^2 \over r'^2} \right) 2Gh(r') e^{-\int_{r'}^r \! dr'' {2Gh(r'')\over r''}}~.  }
Writing $ 1+{GQ^2 \over r'^2} = f(r')+{2GM\over r'} $, and observing that when inserted into \rf{d11vz} the  ${2GM\over r'} $ contribution yields a total $r'$-derivative, we arrive at  \rf{d11v}. 
%
%
%
%

\bibliographystyle{bibstyle2017}
\bibliography{collection}

\providecommand{\href}[2]{#2}\begingroup\begin{thebibliography}{10}

\bibitem{Preskill:1991tb}
J.~Preskill, P.~Schwarz, A.~D. Shapere, S.~Trivedi, and F.~Wilczek, {\it {Limitations on the statistical description of black holes}},  \href{http://dx.doi.org/10.1142/S0217732391002773}{{\sf Mod. Phys. Lett. A} {\sf {6} }{\sf (1991) }{\sf 2353--2362}}.

\bibitem{Almheiri:2014cka}
A.~Almheiri and J.~Polchinski, {\it {Models of AdS$_{2}$ backreaction and holography}},  \href{http://dx.doi.org/10.1007/JHEP11(2015)014}{{\sf JHEP} {\sf {11} }{\sf (2015) }{\sf 014}}, \href{http://arxiv.org/abs/1402.6334}{{\ttfamily arXiv:1402.6334 [hep-th]}}.

\bibitem{Jackiw:1984je}
R.~Jackiw, {\it {Lower Dimensional Gravity}},  \href{http://dx.doi.org/10.1016/0550-3213(85)90448-1}{{\sf Nucl. Phys. B} {\sf {252} }{\sf (1985) }{\sf 343--356}}.

\bibitem{Teitelboim:1983ux}
C.~Teitelboim, {\it {Gravitation and Hamiltonian Structure in Two Space-Time Dimensions}},  \href{http://dx.doi.org/10.1016/0370-2693(83)90012-6}{{\sf Phys. Lett. B} {\sf {126} }{\sf (1983) }{\sf 41--45}}.

\bibitem{Jensen:2016pah}
K.~Jensen, {\it {Chaos in AdS$_2$ Holography}},  \href{http://dx.doi.org/10.1103/PhysRevLett.117.111601}{{\sf Phys. Rev. Lett.} {\sf {117} }{\sf no.~11, }{\sf (2016) }{\sf 111601}}, \href{http://arxiv.org/abs/1605.06098}{{\ttfamily arXiv:1605.06098 [hep-th]}}.

\bibitem{Maldacena:2016upp}
J.~Maldacena, D.~Stanford, and Z.~Yang, {\it {Conformal symmetry and its breaking in two dimensional Nearly Anti-de-Sitter space}},  \href{http://dx.doi.org/10.1093/ptep/ptw124}{{\sf PTEP} {\sf {2016} }{\sf no.~12, }{\sf (2016) }{\sf 12C104}}, \href{http://arxiv.org/abs/1606.01857}{{\ttfamily arXiv:1606.01857 [hep-th]}}.

\bibitem{Engelsoy:2016xyb}
J.~Engels{\"o}y, T.~G. Mertens, and H.~Verlinde, {\it {An investigation of AdS$_{2}$ backreaction and holography}},  \href{http://dx.doi.org/10.1007/JHEP07(2016)139}{{\sf JHEP} {\sf {07} }{\sf (2016) }{\sf 139}}, \href{http://arxiv.org/abs/1606.03438}{{\ttfamily arXiv:1606.03438 [hep-th]}}.

\bibitem{Stanford:2017thb}
D.~Stanford and E.~Witten, {\it {Fermionic Localization of the Schwarzian Theory}},  \href{http://dx.doi.org/10.1007/JHEP10(2017)008}{{\sf JHEP} {\sf {10} }{\sf (2017) }{\sf 008}}, \href{http://arxiv.org/abs/1703.04612}{{\ttfamily arXiv:1703.04612 [hep-th]}}.

\bibitem{Iliesiu:2020qvm}
L.~V. Iliesiu and G.~J. Turiaci, {\it {The statistical mechanics of near-extremal black holes}},  \href{http://dx.doi.org/10.1007/JHEP05(2021)145}{{\sf JHEP} {\sf {05} }{\sf (2021) }{\sf 145}}, \href{http://arxiv.org/abs/2003.02860}{{\ttfamily arXiv:2003.02860 [hep-th]}}.

\bibitem{Heydeman:2020hhw}
M.~Heydeman, L.~V. Iliesiu, G.~J. Turiaci, and W.~Zhao, {\it {The statistical mechanics of near-BPS black holes}},  \href{http://dx.doi.org/10.1088/1751-8121/ac3be9}{{\sf J. Phys. A} {\sf {55} }{\sf no.~1, }{\sf (2022) }{\sf 014004}}, \href{http://arxiv.org/abs/2011.01953}{{\ttfamily arXiv:2011.01953 [hep-th]}}.

\bibitem{Mertens:2022irh}
T.~G. Mertens and G.~J. Turiaci, {\it {Solvable models of quantum black holes: a review on Jackiw{\textendash}Teitelboim gravity}},  \href{http://dx.doi.org/10.1007/s41114-023-00046-1}{{\sf Living Rev. Rel.} {\sf {26} }{\sf no.~1, }{\sf (2023) }{\sf 4}}, \href{http://arxiv.org/abs/2210.10846}{{\ttfamily arXiv:2210.10846 [hep-th]}}.

\bibitem{Almheiri:2016fws}
A.~Almheiri and B.~Kang, {\it {Conformal Symmetry Breaking and Thermodynamics of Near-Extremal Black Holes}},  \href{http://dx.doi.org/10.1007/JHEP10(2016)052}{{\sf JHEP} {\sf {10} }{\sf (2016) }{\sf 052}}, \href{http://arxiv.org/abs/1606.04108}{{\ttfamily arXiv:1606.04108 [hep-th]}}.

\bibitem{Hawking:1975vcx}
S.~W. Hawking, {\it {Particle Creation by Black Holes}},  \href{http://dx.doi.org/10.1007/BF02345020}{{\sf Commun. Math. Phys.} {\sf {43} }{\sf (1975) }{\sf 199--220}}. [Erratum: Commun.Math.Phys. 46, 206 (1976)].

\bibitem{Kolanowski:2024zrq}
M.~Kolanowski, D.~Marolf, I.~Rakic, M.~Rangamani, and G.~J. Turiaci, {\it {Looking at extremal black holes from very far away}},  \href{http://dx.doi.org/10.1007/JHEP04(2025)020}{{\sf JHEP} {\sf {04} }{\sf (2025) }{\sf 020}}, \href{http://arxiv.org/abs/2409.16248}{{\ttfamily arXiv:2409.16248 [hep-th]}}.

\bibitem{Nayak:2018qej}
P.~Nayak, A.~Shukla, R.~M. Soni, S.~P. Trivedi, and V.~Vishal, {\it {On the Dynamics of Near-Extremal Black Holes}},  \href{http://dx.doi.org/10.1007/JHEP09(2018)048}{{\sf JHEP} {\sf {09} }{\sf (2018) }{\sf 048}}, \href{http://arxiv.org/abs/1802.09547}{{\ttfamily arXiv:1802.09547 [hep-th]}}.

\bibitem{Moitra:2018jqs}
U.~Moitra, S.~P. Trivedi, and V.~Vishal, {\it {Extremal and near-extremal black holes and near-CFT$_{1}$}},  \href{http://dx.doi.org/10.1007/JHEP07(2019)055}{{\sf JHEP} {\sf {07} }{\sf (2019) }{\sf 055}}, \href{http://arxiv.org/abs/1808.08239}{{\ttfamily arXiv:1808.08239 [hep-th]}}.

\bibitem{Brown:2024ajk}
A.~R. Brown, L.~V. Iliesiu, G.~Penington, and M.~Usatyuk, {\it {The evaporation of charged black holes}},  \href{http://arxiv.org/abs/2411.03447}{{\ttfamily arXiv:2411.03447 [hep-th]}}.

\bibitem{Lin:2025wof}
G.~Lin, L.~V. Iliesiu, and M.~Usatyuk, {\it {The evaporation of black holes in supergravity}},  \href{http://arxiv.org/abs/2504.21077}{{\ttfamily arXiv:2504.21077 [hep-th]}}.

\bibitem{Emparan:2025sao}
R.~Emparan, {\it {Quantum cross-section of near-extremal black holes}},  \href{http://dx.doi.org/10.1007/JHEP04(2025)122}{{\sf JHEP} {\sf {04} }{\sf (2025) }{\sf 122}}, \href{http://arxiv.org/abs/2501.17470}{{\ttfamily arXiv:2501.17470 [hep-th]}}.

\bibitem{Biggs:2025nzs}
A.~Biggs, {\it {Following the state of an evaporating charged black hole into the quantum gravity regime}},  \href{http://arxiv.org/abs/2503.02051}{{\ttfamily arXiv:2503.02051 [hep-th]}}.

\bibitem{Emparan:2025qqf}
R.~Emparan and S.~Trezzi, {\it {Quantum Transparency of Near-extremal Black Holes}},  \href{http://arxiv.org/abs/2507.03398}{{\ttfamily arXiv:2507.03398 [hep-th]}}.

\bibitem{Betzios:2025sct}
P.~Betzios, O.~Papadoulaki, and Y.~Zhou, {\it {Near-extremal quantum cross-section for charged fields and superradiance}},  \href{http://arxiv.org/abs/2507.13896}{{\ttfamily arXiv:2507.13896 [hep-th]}}.

\bibitem{Berger:1972pg}
B.~K. Berger, D.~M. Chitre, V.~E. Moncrief, and Y.~Nutku, {\it {Hamiltonian formulation of spherically symmetric gravitational fields}},  \href{http://dx.doi.org/10.1103/PhysRevD.5.2467}{{\sf Phys. Rev. D} {\sf {5} }{\sf (1972) }{\sf 2467--2470}}.

\bibitem{Unruh:1976db}
W.~G. Unruh, {\it {Notes on black hole evaporation}},  \href{http://dx.doi.org/10.1103/PhysRevD.14.870}{{\sf Phys. Rev. D} {\sf {14} }{\sf (1976) }{\sf 870}}.

\bibitem{Birrell:1982ix}
N.~D. Birrell and P.~C.~W. Davies, \href{http://dx.doi.org/10.1017/CBO9780511622632}{{\it {Quantum Fields in Curved Space}}, }.
\newblock Cambridge Monographs on Mathematical Physics. Cambridge University Press, Cambridge, UK, 1982.

\bibitem{Fischler:1990pk}
W.~Fischler, D.~Morgan, and J.~Polchinski, {\it {Quantization of False Vacuum Bubbles: A Hamiltonian Treatment of Gravitational Tunneling}},  \href{http://dx.doi.org/10.1103/PhysRevD.42.4042}{{\sf Phys. Rev. D} {\sf {42} }{\sf (1990) }{\sf 4042--4055}}.

\bibitem{Husain:2005gx}
V.~Husain and O.~Winkler, {\it {Flat slice Hamiltonian formalism for dynamical black holes}},  \href{http://dx.doi.org/10.1103/PhysRevD.71.104001}{{\sf Phys. Rev. D} {\sf {71} }{\sf (2005) }{\sf 104001}}, \href{http://arxiv.org/abs/gr-qc/0503031}{{\ttfamily arXiv:gr-qc/0503031}}.

\bibitem{Regge:1974zd}
T.~Regge and C.~Teitelboim, {\it {Role of Surface Integrals in the Hamiltonian Formulation of General Relativity}},  \href{http://dx.doi.org/10.1016/0003-4916(74)90404-7}{{\sf Annals Phys.} {\sf {88} }{\sf (1974) }{\sf 286}}.

\bibitem{Stephens:1993an}
C.~R. Stephens, G.~'t~Hooft, and B.~F. Whiting, {\it {Black hole evaporation without information loss}},  \href{http://dx.doi.org/10.1088/0264-9381/11/3/014}{{\sf Class. Quant. Grav.} {\sf {11} }{\sf (1994) }{\sf 621--648}}, \href{http://arxiv.org/abs/gr-qc/9310006}{{\ttfamily arXiv:gr-qc/9310006}}.

\bibitem{Kiem:1995iy}
Y.~Kiem, H.~L. Verlinde, and E.~P. Verlinde, {\it {Black hole horizons and complementarity}},  \href{http://dx.doi.org/10.1103/PhysRevD.52.7053}{{\sf Phys. Rev. D} {\sf {52} }{\sf (1995) }{\sf 7053--7065}}, \href{http://arxiv.org/abs/hep-th/9502074}{{\ttfamily arXiv:hep-th/9502074}}.

\bibitem{Shenker:2013pqa}
S.~H. Shenker and D.~Stanford, {\it {Black holes and the butterfly effect}},  \href{http://dx.doi.org/10.1007/JHEP03(2014)067}{{\sf JHEP} {\sf {03} }{\sf (2014) }{\sf 067}}, \href{http://arxiv.org/abs/1306.0622}{{\ttfamily arXiv:1306.0622 [hep-th]}}.

\bibitem{Polchinski:2015cea}
J.~Polchinski, {\it {Chaos in the black hole S-matrix}},  \href{http://arxiv.org/abs/1505.08108}{{\ttfamily arXiv:1505.08108 [hep-th]}}.

\bibitem{Kraus:1994by}
P.~Kraus and F.~Wilczek, {\it {Selfinteraction correction to black hole radiance}},  \href{http://dx.doi.org/10.1016/0550-3213(94)00411-7}{{\sf Nucl. Phys. B} {\sf {433} }{\sf (1995) }{\sf 403--420}}, \href{http://arxiv.org/abs/gr-qc/9408003}{{\ttfamily arXiv:gr-qc/9408003}}.

\bibitem{Kraus:1994fj}
P.~Kraus and F.~Wilczek, {\it {Effect of selfinteraction on charged black hole radiance}},  \href{http://dx.doi.org/10.1016/0550-3213(94)00588-6}{{\sf Nucl. Phys. B} {\sf {437} }{\sf (1995) }{\sf 231--242}}, \href{http://arxiv.org/abs/hep-th/9411219}{{\ttfamily arXiv:hep-th/9411219}}.

\bibitem{Keski-Vakkuri:1996wom}
E.~Keski-Vakkuri and P.~Kraus, {\it {Microcanonical D-branes and back reaction}},  \href{http://dx.doi.org/10.1016/S0550-3213(97)00085-0}{{\sf Nucl. Phys. B} {\sf {491} }{\sf (1997) }{\sf 249--262}}, \href{http://arxiv.org/abs/hep-th/9610045}{{\ttfamily arXiv:hep-th/9610045}}.

\bibitem{Balbinot:2023vcm}
R.~Balbinot and A.~Fabbri, {\it {The Unruh Vacuum and the {\textquotedblleft}In-Vacuum{\textquotedblright} in Reissner-Nordstr{\"o}m Spacetime {\textdagger}}},  \href{http://dx.doi.org/10.3390/universe10010018}{{\sf Universe} {\sf {10} }{\sf no.~1, }{\sf (2024) }{\sf 18}}, \href{http://arxiv.org/abs/2311.09943}{{\ttfamily arXiv:2311.09943 [gr-qc]}}.

\end{thebibliography}\endgroup

\end{document}